\documentclass[a4paper,11pt]{article}
\pdfoutput=1
\usepackage{graphicx,rotating,hyperref,slashed,amsmath,xcolor,amssymb,amsfonts,colortbl,cite, subfigure,float}
\makeatletter 
\usepackage{amsmath}
\usepackage{graphics} 
\usepackage{graphicx} 
\graphicspath{{plots/}}
 
\hypersetup{colorlinks,bookmarksopen,bookmarksnumbered,
linkcolor=blus,pdfstartview=FitH,urlcolor=rossos,citecolor=verde}
\numberwithin{equation}{section}

\hypersetup{%
    ,urlcolor=rossos
    ,citecolor=rossos
    ,linkcolor=rossos
    }

\AtBeginDocument{
  \hypersetup{
    urlcolor=verdes,
    citecolor=verdes,
    linkcolor=verdes,
  }%
}

\allowdisplaybreaks

\def\lsim{\mathrel{\rlap{\lower3pt\hbox{\hskip0pt$\sim$}}
   \raise1pt\hbox{$<$}}}         
\def\gsim{\mathrel{\rlap{\lower4pt\hbox{\hskip1pt$\sim$}}
   \raise1pt\hbox{$>$}}}         

 \newcommand{\sfootnote}[1]{} 
\definecolor{bluc}{cmyk}{1,1,0,0.1}
\definecolor{rossoCP3}{cmyk}{0,.88,.77,.40}
\definecolor{rosso}{cmyk}{0,1,1,0.4}
\definecolor{rossos}{cmyk}{0,1,1,0.55}
\definecolor{rossoc}{cmyk}{0,1,1,0.2}
\definecolor{verdes}{cmyk}{0.92,0,0.59,0.4}

\hypersetup{colorlinks, bookmarksopen, bookmarksnumbered,
citecolor=verdes, linkcolor=bluc, pdfstartview=FitH, urlcolor=rossos}

\newcommand{\mio}[1]{}

\definecolor{Gray}{gray}{0.95}

\usepackage{multicol}
\usepackage{color}
\definecolor{rosso}{cmyk}{0,1,1,0.4}
\definecolor{rossos}{cmyk}{0,1,1,0.55}
\definecolor{rossoc}{cmyk}{0,1,1,0.2}
\definecolor{blu}{cmyk}{1,1,0,0.3}
\definecolor{blus}{cmyk}{1,1,0,0.6}
\definecolor{bluc}{cmyk}{1,1,0,0.1}
\definecolor{verde}{cmyk}{0.92,0,0.59,0.25}
\definecolor{verdec}{cmyk}{0.92,0,0.59,0.15}
\definecolor{verdes}{cmyk}{0.92,0,0.59,0.4}

\def\circa#1{\,\raise.3ex\hbox{$#1$\kern-.75em\lower1ex\hbox{$\sim$}}\,}

\newcommand{\beq}{\begin{equation}}
\newcommand{\eeq}{\end{equation}}

\newcommand{\bea}{\begin{eqnarray}}
\newcommand{\eea}{\end{eqnarray}}
\newcommand{\be}{\begin{equation}}
\newcommand{\ee}{\end{equation}}
\newfam\rsfsfam
\def\mathscr#1{{\fam\rsfsfam\relax#1}}

\def\circa#1{\,\raise.3ex\hbox{$#1$\kern-.75em\lower1ex\hbox{$\sim$}}\,}
\makeatletter

\def\hhref#1{\href{http://arxiv.org/abs/#1}{arXiv:#1}} 

\newcommand{\doi}[1]{\href{http://dx.doi.org/#1}{[doi]}}

\def\hhref#1{\href{http://arxiv.org/abs/#1}{arXiv:#1}} 
 
\def\art{\@ifnextchar[{\eart}{\oart}}
\def\eart[#1]#2#3#4#5#6{{\rm #2}, {\em #3 \bf #4} {\rm (#6) #5} ({\em #1})}

\def\article{\@ifnextchar[{\earticle}{\oarticle}}
\def\oarticle#1#2#3#4#5#6{{\rm #1}, {\em ``#6''}, {\rm #2 #3 (#5) #4}}
\def\earticle[#1]#2#3#4#5#6#7{{\rm #2}, {\em ``#7''}, {\rm #3 #4 (#6) #5}  [\hhref{#1}]}
\def\hepart[#1]#2{{\rm #2, \em#1}}
\def\heparticle[#1]#2#3{#2, {\em ``#3''} [\hhref{#1}]}

\newcounter{alphaequation}[equation]
\def\thealphaequation{\theequation\hbox to
0.6em{\hfil\alph{alphaequation}\hfil}}
\def\eqnsystem#1{
\def\@eqnnum{{\rm (\thealphaequation)}}
\def\@@eqncr{\let\@tempa\relax \ifcase\@eqcnt \def\@tempa{& & &} \or
  \def\@tempa{& &}\or \def\@tempa{&}\fi\@tempa
  \if@eqnsw\@eqnnum\refstepcounter{alphaequation}\fi
\global\@eqnswtrue\global\@eqcnt=0\cr}
\refstepcounter{equation} \let\@currentlabel\theequation \def\@tempb{#1}
\ifx\@tempb\empty\else\label{#1}\fi
\refstepcounter{alphaequation}
\let\@currentlabel\thealphaequation
\global\@eqnswtrue\global\@eqcnt=0 \tabskip\@centering\let\\=\@eqncr
$$\halign to \displaywidth\bgroup \@eqnsel\hskip\@centering
$\displaystyle\tabskip\z@{##}$&\global\@eqcnt\@ne
\hskip2\arraycolsep\hfil${##}$\hfil& \global\@eqcnt\tw@\hskip2\arraycolsep
$\displaystyle\tabskip\z@{##}$\hfil
\tabskip\@centering&\llap{##}\tabskip\z@\cr}
\def\endeqnsystem{\@@eqncr\egroup$$\global\@ignoretrue} \makeatother

\definecolor{fiorentina}{rgb}{.5,0,.5}

\begin{document}

\vspace{1truecm}
 \begin{center}
\boldmath

{\textbf{\LARGE  
{ 
Ultralight  dark matter from non-slowroll inflation}}
}
\unboldmath
\end{center}
\unboldmath

\vspace{-0.2cm}

\begin{center}
\vspace{0.1truecm}

\renewcommand{\thefootnote}{\fnsymbol{footnote}}
\begin{center} 
{\fontsize{13}{30}
\selectfont  
 Martina La Rosa$^{a,b\!}$ \footnote{\texttt{martinalarosa20@gmail.com}},  Gianmassimo Tasinato$^{a,b\!}$ \footnote{\texttt{g.tasinato2208.at.gmail.com}}
} 
\end{center}


\begin{center}

\vskip 6pt
\textsl{$^a$ Physics Department, Swansea University, SA2 8PP, UK}
\\
\textsl{$^{b}$ Dipartimento di Fisica e Astronomia, Universit\`a di Bologna,\\
 INFN, Sezione di Bologna,  viale B. Pichat 6/2, 40127 Bologna,   Italy}
\vskip 4pt

\end{center}
\hskip0.1cm

\begin{abstract}
\noindent
The longitudinal mode of a massive vector field, generated  during inflation, offers a  well-motivated and phenomenologically rich candidate for dark matter. We show that a rapid variation in the mass of the  vector boson, occurring during a brief phase of non-slowroll inflationary evolution, can naturally give rise to extremely small vector masses after inflation ends, corresponding to an ultralight dark matter candidate.
This mechanism predicts a stochastic gravitational-wave background, generated at second order by non-adiabatic longitudinal vector fluctuations and amplified at very low frequencies, yielding a distinctive observational signature of the scenario. By leveraging a brief departure from slowroll dynamics during inflation -- commonly invoked in scenarios that produce primordial black holes -- our framework establishes a novel connection between ultralight vector dark matter and primordial black hole physics, suggesting  a possible unified setting for mixed dark matter scenarios.
\end{abstract}

\end{center}

\renewcommand{\thefootnote}{\arabic{footnote}}
\setcounter{footnote}{0}

\section{Introduction}

Determining the nature of dark matter (DM) is one of the most important open 
problems in modern physics. Its feeble interactions with Standard
Model particles, if any besides gravity,  make its detection extremely
challenging.  See, for example,~\cite{Bertone:2004pz,Feng:2010gw} for reviews.

In this work, building on the analysis of~\cite{Marriott-Best:2025sez}, we explore a variation of the scenario originally proposed in~\cite{Graham:2015rva}, in which DM consists of the longitudinal modes of a massive dark photon produced purely through gravitational effects during inflation. The setup of \cite{Graham:2015rva} is particularly minimal, requiring only a single massive vector field as the dark matter candidate, which is efficiently generated in the very early Universe.
We investigate how this framework is modified by the presence of a brief phase of non-slowroll evolution during inflation, which causes a rapid variation of the vector mass. Our extension of the original scenario allows for the generation of dark matter with extremely small masses, on the order of $m = 10^{-21}$ eV or smaller. A key prediction of this mechanism is the simultaneous production of a stochastic background of gravitational waves, peaking at very low frequencies, which offers a distinctive smoking gun signature for the scenario.
By introducing a short departure from slowroll inflation through carefully chosen initial conditions, our approach provides a new connection between the framework of~\cite{Graham:2015rva} and the class of models associated with the generation of primordial black holes (see, e.g.,~\cite{Ozsoy:2023ryl} for a review). This opens up an intriguing link between ultralight vector dark matter and primordial black hole physics, offering a new observational probe of inflationary dynamics across a wide range of scales, and suggesting a framework for mixed dark matter
scenarios. 

\smallskip

The gravitational production of scalar fields in inflationary scenarios has been studied in several earlier works, including dark matter production (see for example \cite{Ford:1986sy,Yajnik:1990un,Chung:1998zb}). Ref.~\cite{Graham:2015rva} extended these analyses to the spin-1 case, establishing a compelling connection to DM. In particular, it emphasized the role of the vector  longitudinal component, which arises due to the mass term, and whose Lagrangian in a cosmological  background was first   analyzed in~\cite{Himmetoglu:2008hx}. 
The framework of~\cite{Graham:2015rva} has since been further developed and analyzed in a number of theoretical studies~\cite{Bastero-Gil:2018uel,Ema:2019yrd,Nakai:2020cfw,Ahmed:2020fhc,Kolb:2020fwh,Salehian:2020asa,Moroi:2020has,Arvanitaki:2021qlj,Sato:2022jya,Barman:2021qds,Redi:2022zkt,Ozsoy:2023gnl}, and its phenomenological implications have been investigated in~\cite{An:2014twa,McDermott:2019lch,Lee:2020wfn,Amin:2022pzv,Siemonsen:2022ivj,East:2022ppo,Pierce:2018xmy,Nomura:2019cvc,PPTA:2022eul,Unal:2022ooa,Yu:2023iog}.

\smallskip

 Sections~\ref{sec_setu}--\ref{sec_dmam}
of this work  show that a brief phase of non-slowroll evolution during inflation significantly modifies the frequency profile of the longitudinal vector sector. 
This can occur, for example, when the vector mass depends on the inflaton velocity, which can change significantly over a brief period when slow-roll conditions are violated. 
The resulting spectrum exhibits a characteristic peaked structure, whose amplitude and position depend sensitively on both the vector mass and the parameters controlling the non-slowroll epoch. Notably, this phase alters the infrared behavior of the longitudinal spectrum, leading to a $k^6$ growth -- compared to the $k^2$ scaling found in the original scenario in the infrared part of the spectrum~\cite{Graham:2015rva}.
This modification has important phenomenological implications: it allows for viable dark matter candidates with extremely small vector masses,  thereby providing a concrete realization of ultralight spin-1 dark matter. Moreover, it represents an interesting generalization of previous findings on the maximal growth of the spectrum of adiabatic fluctuations in models of inflation including a brief phase of non-slowroll evolution \cite{Byrnes:2018txb,Ozsoy:2019lyy}.

In Section~\ref{sec_pheno},
building on~\cite{Marriott-Best:2025sez}, we  demonstrate that this scenario naturally gives rise to a stochastic background of gravitational waves, generated at second order in perturbation theory by the enhanced longitudinal vector spectrum at small scales. The amplitude of this gravitational signal increases with decreasing vector mass, and  the peak of the spectrum lies at very low frequencies, around \( 10^{-15} - 10^{-13} \,\mathrm{Hz} \), making detection observationally challenging. Nevertheless, we discuss possible techniques to probe such signals, and we outline additional phenomenological consequences of this framework. A Section of Conclusions and two technical Appendixes conclude the article. Natural units are adopted throughout.

\section{The setup}
\label{sec_setu}

We present a cosmological scenario involving a massive vector field whose mass evolves dynamically during inflation. The nonzero vector mass explicitly breaks an Abelian gauge symmetry, thereby rendering the longitudinal mode of the vector field physical and dynamical. If the mass undergoes a rapid variation over a brief interval during inflation, to then stabilize to a constant value by when inflation ends,   the system exhibits distinctive features, which we will analyse in detail. We argue that this mechanism provides a compelling framework for generating dark matter in the form of the longitudinal mode of a massive dark vector field, using ideas borrowed from physics
of primordial black hole formation, and with interesting phenomenological ramifications.
The longitudinal vector dark matter
 abundance and  its properties are controlled by the vector mass scale and its time variation during inflation. We show that this scenario allows for a new mechanism
 to produce ultralight dark matter in the form of vector bosons during inflation. 
 This work offers an explicit realization of the phenomenological
 ideas developed in \cite{Marriott-Best:2025sez}.

\smallskip
We consider a generalization of the covariant vector-tensor action studied, for instance, in~\cite{Demozzi:2009fu}, by introducing a spacetime-dependent mass term in a cosmological context:
\be
\label{eq_dia1}
S = \int d^4 x\, \sqrt{-g} \left[
\frac{R}{2} - \frac{1}{4} F_{\mu\nu} F^{\mu\nu} - \frac{M^2}{2} A_\mu A^\mu
\right]\,,
\ee
with $F_{\mu\nu}=\partial A_\nu-\partial_\nu A_\mu$. 
The quantity $M^2$ is a mass parameter that can depend on spacetime coordinates -- for example, through its coupling to dynamical fields active during inflation, such as the inflaton. We focus on a spatially flat, homogeneous and isotropic background metric:
\be
ds^2 = a^2(\tau) \left[ 
- d\tau^2 + \delta_{ij}\, dx^i dx^j
\right]\,.
\ee

We assume that the vector field $A_\mu$ has no homogeneous background value and instead behaves as a perturbation propagating on the cosmological background. Its decomposition takes the form:
\be\label{eq_vede}
A_\mu(x) = \left( A_0(x),\, \partial_i \varphi(x) + A_i^T(x) \right)\,,
\ee
where $A_i^T$ is transverse, satisfying $\partial_i A_i^T = 0$.  
In Eq.~\eqref{eq_dia1}, we choose the mass term to take the form:
\be
\label{def_capm}
M^2(\tau) = \frac{m^2\, J^2(\tau)}{a^2(\tau)}\,,
\ee
where \( m \) is a constant mass scale, corresponding to the physical vector mass at the end of inflation $\tau_R$ , and \( J(\tau) \) is a dimensionless function of conformal time. For simplicity, we impose the condition \( J(\tau_R) = a(\tau_R) \), so that \( M(\tau_R) = m \). The time dependence of \( J(\tau) \) can be naturally motivated if it originates from a coupling to the inflaton field, whose dynamics during inflation govern the evolution of the effective vector mass. 
 In our discussion 
we are guided by scenarios in which the inflaton velocity undergoes a brief, rapid variation --such as in ultra-slowroll~\cite{Kinney:2005vj,Martin:2012pe,Dimopoulos:2017ged} or constant-roll~\cite{Motohashi:2014ppa,Inoue:2001zt,Tzirakis:2007bf} inflation (see e.g.~\cite{Ozsoy:2023ryl} for a review).  These rapid transitions in the inflaton's velocity are expected to manifest as a sharp, short-duration variation in the function \( J(\tau) \).
Although we do not specify a particular inflationary scenario for our arguments, we develop in Appendix \ref{app_mod} a representative
inflation model with some of the characteristics outlined above.
Building on these hypothesis,
the central problem we address in this work is how to analyze the dynamics of vector fluctuations when \( J(\tau) \) exhibits such a rapid transition during inflation.  Succeeding on this aim will us
to enlarge the parameter space of the longitudinal vector dark matter model \cite{Graham:2015rva} allowing for small dark matter masses with interesting
phenomenological consequences.

\smallskip

In what follows, we neglect fluctuations of the metric, assuming that their 
contributions do not spoil the early universe amplification mechanism of \cite{Graham:2015rva}. Hence we focus on
 the quadratic action for vector fluctuations around a homogeneous and isotropic background, assuming the vector field has no homogeneous vacuum expectation value, reads:
\be
\label{eq_dia}
S = \int d\tau\, d^3x\, a^2(\tau)
\left[ -\frac{1}{4} F_{\mu\nu} F^{\mu\nu} - \frac{m^2 J^2(\tau)}{2 a^2(\tau)} A_\mu A^\mu \right]\,.
\ee

As described in the decomposition~\eqref{eq_vede}, the gauge field contains transverse vector modes and a scalar (longitudinal) degree of freedom. In what follows, we focus exclusively on the scalar sector, described by the fields $A_0$ and $\varphi$. In Fourier space, the physical longitudinal mode of the vector field is given by:
\be
A_{L\,\mathbf{k}}(\tau) = i k\, \varphi_{\mathbf{k}}(\tau)\,.
\ee

Solving the equation of motion for the nondynamical component $A_0$ yields the following relation in Fourier space:
\be
\label{rel_aze}
A_{0\,\mathbf{k}}(\tau) = \frac{k^2}{k^2 + m^2 J^2(\tau)}\, \varphi'_{\mathbf{k}}(\tau)
= \frac{-i k}{k^2 + m^2 J^2(\tau)}\, A'_{L\,\mathbf{k}}(\tau)\,.
\ee

Substituting Eq.~\eqref{rel_aze} into the action, we obtain the following effective quadratic action for the longitudinal scalar mode $\varphi_{\mathbf{k}}$:
\be
\label{eq_dia2}
S = \int d\tau\, d^3k\, \frac{k^2 J^2(\tau)}{2} \left[
\frac{m^2}{k^2 + m^2 J^2(\tau)}\, \varphi'_{\mathbf{k}}(\tau) \varphi'_{-\mathbf{k}} (\tau)
- m^2\, \varphi_{\mathbf{k}} (\tau) \varphi_{-\mathbf{k}}(\tau)
\right]\,.
\ee
To bring the kinetic term into canonical form, we define the canonically normalized field:
\be
\label{pi-can}
\pi_{\mathbf{k}}(\tau) \equiv \frac{k m J(\tau)}{\sqrt{k^2 + m^2 J^2(\tau)}}\, \varphi_{\mathbf{k}}(\tau)\,.
\ee
In terms of $\pi_{\mathbf{k}}$, the quadratic action becomes (we understand here
the time dependence of the various quantities):
\begin{equation}
\label{ac-can}
S = \frac{1}{2} \int d\tau\, d^3k \left[
\pi'_{\mathbf{k}}\, \pi'_{-\mathbf{k}} - 
\left( k^2 + m^2 J^2 + \frac{3 k^2 m^2 J'^2}{(k^2 + m^2 J^2)^2} 
- \frac{k^2}{k^2 + m^2 J^2}\, \frac{J''}{J} \right)
\pi_{\mathbf{k}}\, \pi_{-\mathbf{k}}
\right]\,,
\end{equation}
which constitutes the starting point for our discussion. In what follows, we analytically solve the 
corresponding evolution equations in different regimes -- inflation and radiation domination -- and demonstrate that vector longitudinal modes can constitute an interesting
dark matter candidate, whose properties depend
on the vector mass $m$ and the characteristics
of the time-dependent function $J(\tau)$. 
 
 \section{Cosmological evolution of the vector longitudinal mode}
 \label{sec_genev}

In this section we analyze the rich cosmological evolution of the longitudinal
mode  of the massive vector field $A_\mu$ in a scenario where inflation
is followed by a radiation domination universe, and
the vector mass $M$ of Eq.~\eqref{def_capm}
changes rapidly during a short period within the inflationary phase.

\subsection{Evolution during inflation}
\label{sec_inf}

We consider the regime in which the physical vector mass is much smaller than the Hubble scale during inflation, $m \ll H_I$, with $m$ the constant mass parameter introduced in Eq.~\ref{def_capm}, and $H_I$
the inflationary Hubble scale, assumed nearly constant during inflation. In fact, throughout this section we assume  a quasi-de Sitter background with scale factor $a(\tau) = -1/(H_I \tau)$. We work under the following hierarchy of scales: at subhorizon scales, the physical momentum satisfies
\bea
\frac{k}{a H_I} \gg 1 \gg \frac{J m}{a H_I}\,,
\eea
while at superhorizon scales we have
\bea
\frac{J m}{a H_I} \ll \frac{k}{a H_I} \ll 1\,.
\eea
Under these conditions, the dominant contributions to the effective action \eqref{ac-can} simplify considerably. In particular, the equation of motion for the canonically normalized scalar field $\pi_{\bf k}$ reduces to
\be
\label{eq_tosol}
\pi_{\bf k}'' (\tau) + \left( k^2 - \frac{J''(\tau)}{J(\tau)} \right) \pi_{\bf k}(\tau) = 0\,.
\ee

Since $\pi_{\bf k} \simeq m\,J(\tau)\,\varphi_{\bf k}$, it follows that the power spectrum of the original scalar mode $\varphi$ is related to that of the canonically normalized mode $\pi$ via
\be
\label{relpphi}
\mathcal{P}_\varphi = \frac{\mathcal{P}_\pi}{m^2 J^2(\tau)}\,.
\ee

To determine the spectrum $\mathcal{P}_\pi$ we follow the approach developed in~\cite{Tasinato:2020vdk,Tasinato:2023ukp} in order 
to deal with Eq.~\eqref{eq_tosol}. We assume that the time-dependent vector mass profile of Eq.~\ref{def_capm} is encoded in the function  
\be
\label{ans_jei}
J(\tau) = a(\tau)\, \sqrt{\omega(\tau)}\,,
\ee
where $\omega(\tau)$ encapsulates the (possibly rapid) time dependence of the physics governing the vector mass during the inflationary epoch. Such a situation can arise, for example, when the vector mass depends on the inflaton velocity, which may vary abruptly over a brief interval in a non-slowroll phase. A concrete illustration of this possibility is provided in Appendix~\ref{app_mod}.

More generally, assuming inflation occurs in the conformal time interval $\tau\le\tau_R$ ($\tau_R$ being
the epoch of the instantaneous reheating) we take
\bea
\label{ans_ome}
\omega(\tau) = 
\begin{cases}
\omega(\tau_1) = \text{const.} & \text{for } \tau < \tau_1\,, \\
\text{smooth but rapidly varying} & \text{for } \tau_1 < \tau < \tau_2\,, \\
1 & \text{for } \tau_2 < \tau_R\,,
\end{cases}
\eea
with the transition interval satisfying $|\tau_1 - \tau_2|/|\tau_1| \ll 1$. This profile
intends to model a brief non-slowroll phase during inflation, treated as nearly instantaneous for analytical tractability, which makes the inflaton velocity rapidly evolve affecting the vector mass. 

\smallskip

We are left with the technical problem to 
 solve Eq.~\eqref{eq_tosol}. We make the Ansatz
\be
\label{ans_pik}
\pi_{\bf k}(\tau) = \frac{e^{-i k \tau} a(\tau) \sqrt{\omega(\tau)} H_I}{\sqrt{2 k^3}}
\left[ 1 + i k \tau + (i k \tau)^2 A_2(\tau) + (i k \tau)^3 A_3(\tau) + \dots \right]\,,
\ee
where the functions $A_n(\tau)$ encode corrections due to the time dependence of $\omega(\tau)$. An arbitrary scale $\tau_0$ can be introduced to keep the series dimensionless; however, our final results are independent of this choice. Plugging this ansatz into Eq.~\eqref{eq_tosol}, we expand in powers of $k$,  and solve order by order by requiring that each term vanishes independently. This method, developed in~\cite{Tasinato:2020vdk}, is well suited to the regime where the non-slowroll phase has infinitesimal duration~\footnote{In realistic scenarios, the phase of non-slowroll evolution is expected to have a finite duration, leading to subleading corrections to the results derived below under the assumption of infinitesimally brief
non-slowroll phase.}, $|\tau_1 - \tau_2|/|\tau_1| \ll 1$. Reference \cite{Tasinato:2020vdk} shows
that each equation can be solved consistently for $A_n(\tau)$ in the limit of brief non-slowroll
phase, and the results can be plugged into \eqref{ans_pik} leading 
to a series which can be resummed analytically. Starting
from this result,  the mode function obtained
within the interval $\tau_1\le\tau\le \tau_2$ can be connected through Israel
junction conditions to the final, slowroll phase of inflation, and evaluated
at the end of the inflationary epoch.

The solution for the mode function $\pi_{\bf k}(\tau)$ depends on a dimensionless parameter $\alpha$, the logarithmic rate of change of $\omega$ at the transition:
\be
\alpha = \left. \frac{d \ln \omega(\tau)}{d \ln \tau} \right|_{\tau = \tau_1}\,,
\ee
which we take positive. Since $d \ln \tau = - d \ln a(\tau)$ during inflation, a positive $\alpha$ corresponds to a negative rate of change for the vector mass $M(\tau)$ during inflation: the vector mass rapidly decreases during
the non-slowroll epoch. This possibility is concretely realized
in the specific model outlined in Appendix \ref{app_mod}.

\smallskip
Formulas simplify significantly in the following idealized limit, motivated by field theory methods  elaborated by  
 't Hooft \cite{tHooft:1973alw}:
\be
\alpha \to \infty\,, \qquad \frac{|\tau_1 - \tau_2|}{|\tau_1|} \to 0\,, \qquad \text{keeping} \quad \frac{\alpha |\tau_1 - \tau_2|}{|\tau_1|} \equiv 2 \Pi_0 \quad \text{finite}\,.
\ee

We refer the reader to  the recent works~\cite{Tasinato:2020vdk,Tasinato:2023ukp,Atkins:2025pvg} for a careful  derivation of these results, where the techniques outlined above are spelled out in all technical details.
In this regime, the resulting power spectrum for the longitudinal mode of the vector
field, when evaluated at the end of inflation, results
\be
\label{NSRpowerspectrum}
\mathcal{P}^{(0)}_{A_L} (k)\,=\, \frac{H_I^2\,k^2}{4 \pi^2 m^2} \, \Pi\left(\frac{k}{k_1}\right)\,,
\ee
where the dimensionless function 
\be
\label{def_cPi}
\Pi(x) = 1 - 4 x \Pi_0 \cos{x}\, j_1(x) + 4 x^2 \Pi_0^2 j_1^2(x)\,
\ee
encodes the imprint of the non-slowroll phase,
which -- as explained above -- we assume is making the vector mass changing rapidly during a small time interval. See in particular \cite{Tasinato:2023ukp} for the work
where the expression \eqref{def_cPi} first appears.
The
quantity $j_1(x)$ in Eq.~\eqref{def_cPi} is the spherical Bessel function of the first kind:
\be
j_1(x) = \frac{\sin x}{x^2} - \frac{\cos x}{x}\,,
\ee
while
the  reference momentum scale $k_1$ appearing in $\Pi(k/k_1)$ is given by
\be
\label{def_kun}
k_1 = H_I\, a(\tau_1) = -\frac{1}{\tau_1}\,,
\ee
and corresponds to the comoving wavenumber of modes exiting the horizon at the onset of the brief non-slowroll phase.

Setting \( \Pi_0 = 0 \) corresponds to a purely slowroll evolution, in which case we recover the standard result~\cite{Graham:2015rva} for Eq.~\eqref{NSRpowerspectrum}, characterized by a \( k^2 \) scaling for the longitudinal modes. 
This mechanism, as described in~\cite{Graham:2015rva}, provides an appealing framework for generating longitudinal vector dark matter during inflation. The resulting blue-tilted spectrum is naturally suppressed at large scales due to the overall \( k^2 \) factor, thereby evading stringent observational bounds on isocurvature perturbations in that regime.

\smallskip

 The phenomenological impact of the non-slowroll phase is illustrated in the upper-left panel of Fig.~\ref{fig_spf1}, and introduces novel features into the dynamics of the system. Activating the parameter \( \Pi_0 \) significantly alters the shape of the longitudinal power spectrum \( \mathcal{P}_{A_L} \), leading to deviations from the standard monotonic \( k^2 \) scaling. In fact, the resulting spectrum exhibits features reminiscent of those seen in scenarios with enhanced curvature perturbations that may trigger primordial black hole (PBH) formation (see~\cite{Ozsoy:2023ryl} for a review).
In particular, an intermediate phase emerges during which the spectrum grows more rapidly, driven by the energy gradient transferred to the longitudinal modes by the time-dependent vector mass. 

Specifically, as shown in Fig.~\ref{fig_spf1}, the power initially rises from large to small scales as \( \mathcal{P}_{A_L} \propto (k/k_1)^2 \), in agreement with the standard behavior~\cite{Graham:2015rva}. This is followed by a dip and a subsequent sharp enhancement, with the spectrum scaling as \( (k/k_1)^6 \) up to momenta around \( k \simeq k_1 \). These modes correspond to scales that exited the horizon during the non-slowroll epoch.
At smaller scales (\( k \gg k_1 \)), the growth rate gradually returns -- on average -- to the milder \( (k/k_1)^2 \) scaling. 
The strong oscillations at small scales arise from the abrupt transition between the non-slowroll phase and the final slowroll epoch during inflation. A smoother matching between the two phases would suppress these oscillations.

The transition from a $k^2$ to a $k^6$ behavior in the infrared region of the spectrum is reminiscent of the maximal power-four enhancement observed in primordial black hole scenarios \cite{Byrnes:2018txb,Ozsoy:2019lyy},
which arise from an initially nearly scale-invariant spectrum. As in that context, we find that the spectrum can acquire at most a fourth-order increase in its momentum dependence relative to its slowroll counterpart \footnote{ It would be interesting to investigate whether the presence of free-streaming particles, or related phenomena, could further affect the shape of the infrared spectrum~\cite{Hook:2020phx}. We leave this question for future work.
}.

The intermediate phase of  \( (k/k_1)^6 \) growth of the longitudinal power spectrum plays a crucial role in determining the final dark matter abundance sourced by the longitudinal vector modes:
in fact, it adds new parameters to the scenario that increase the potentially interesting region of  parameters to explore, with ramifications
for gravitational wave physics.

\subsection{Evolution during radiation domination}
\label{sec_evo}

\begin{figure}[t!]
\centering
    \includegraphics[width=0.49\linewidth]{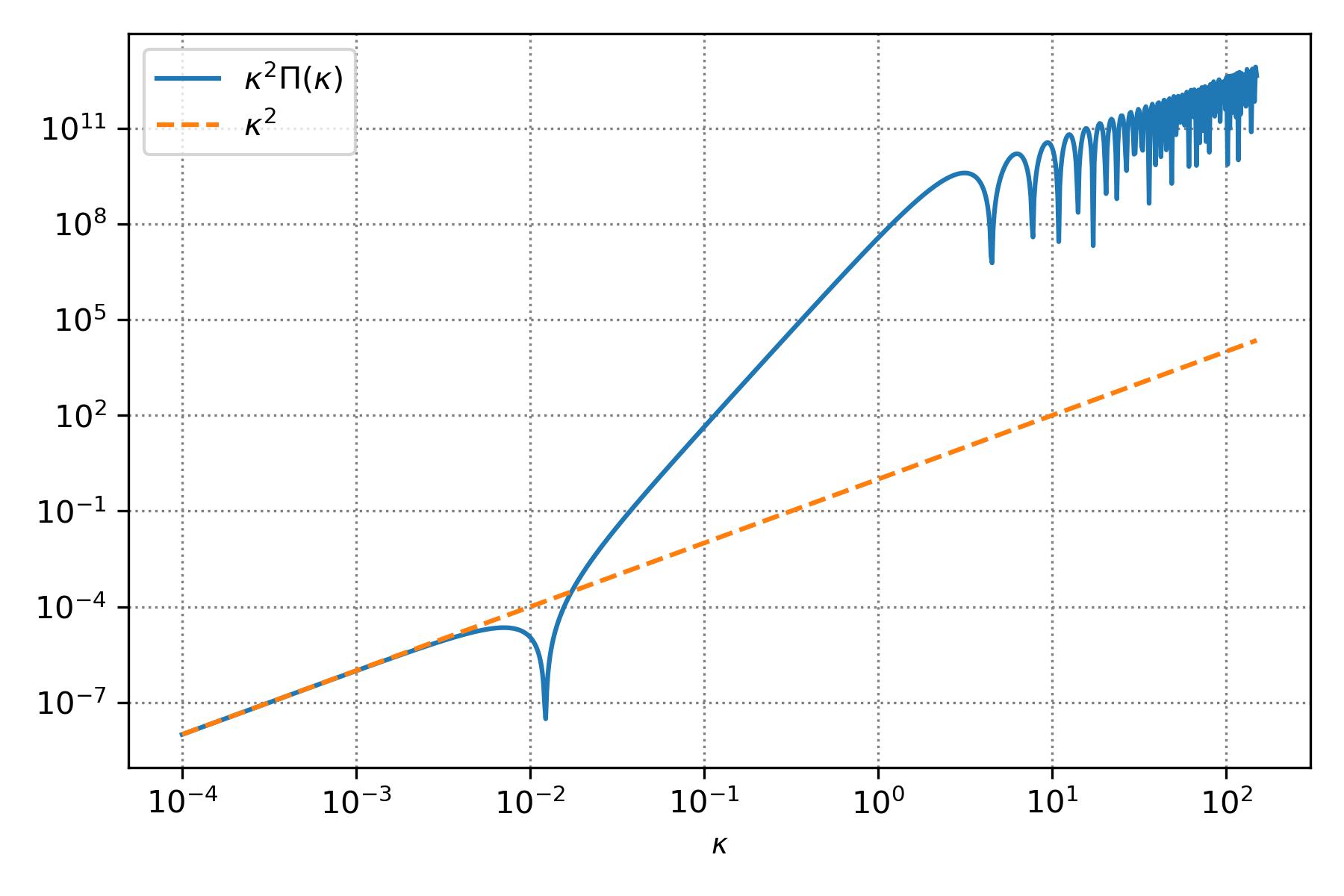}
      \includegraphics[width=0.45\linewidth]{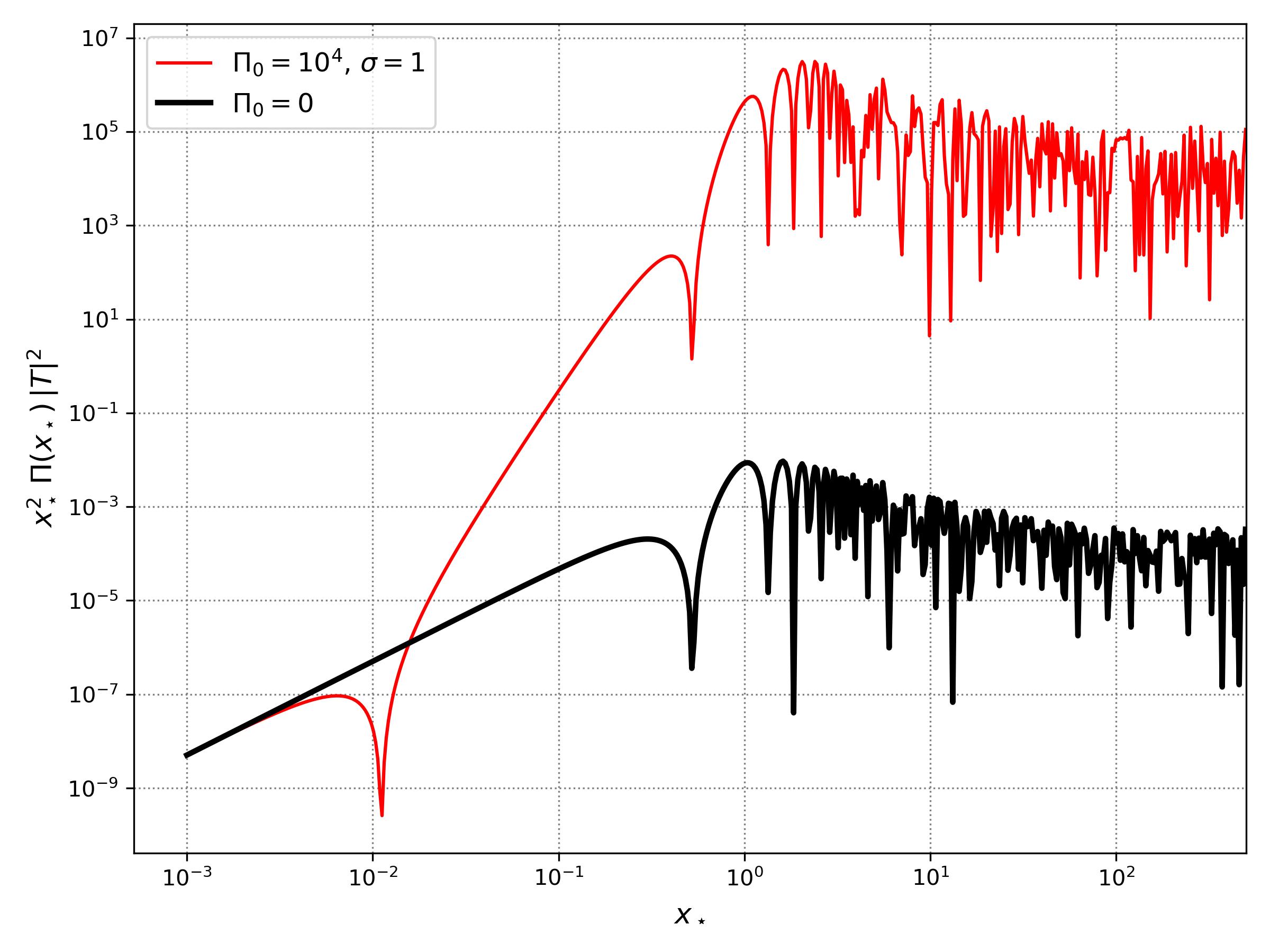}
        \includegraphics[width=0.49\linewidth]{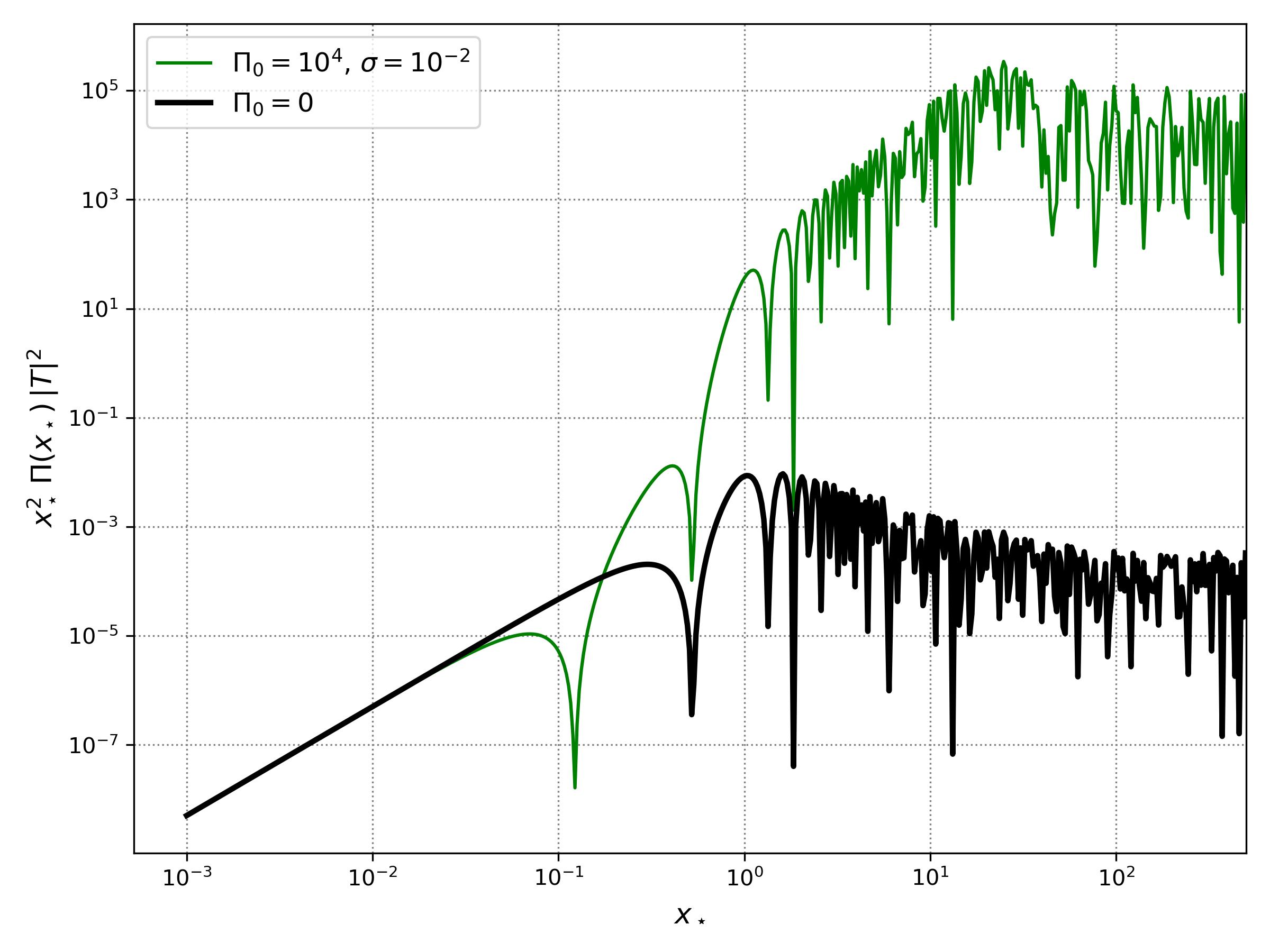}
                \includegraphics[width=0.49\linewidth]{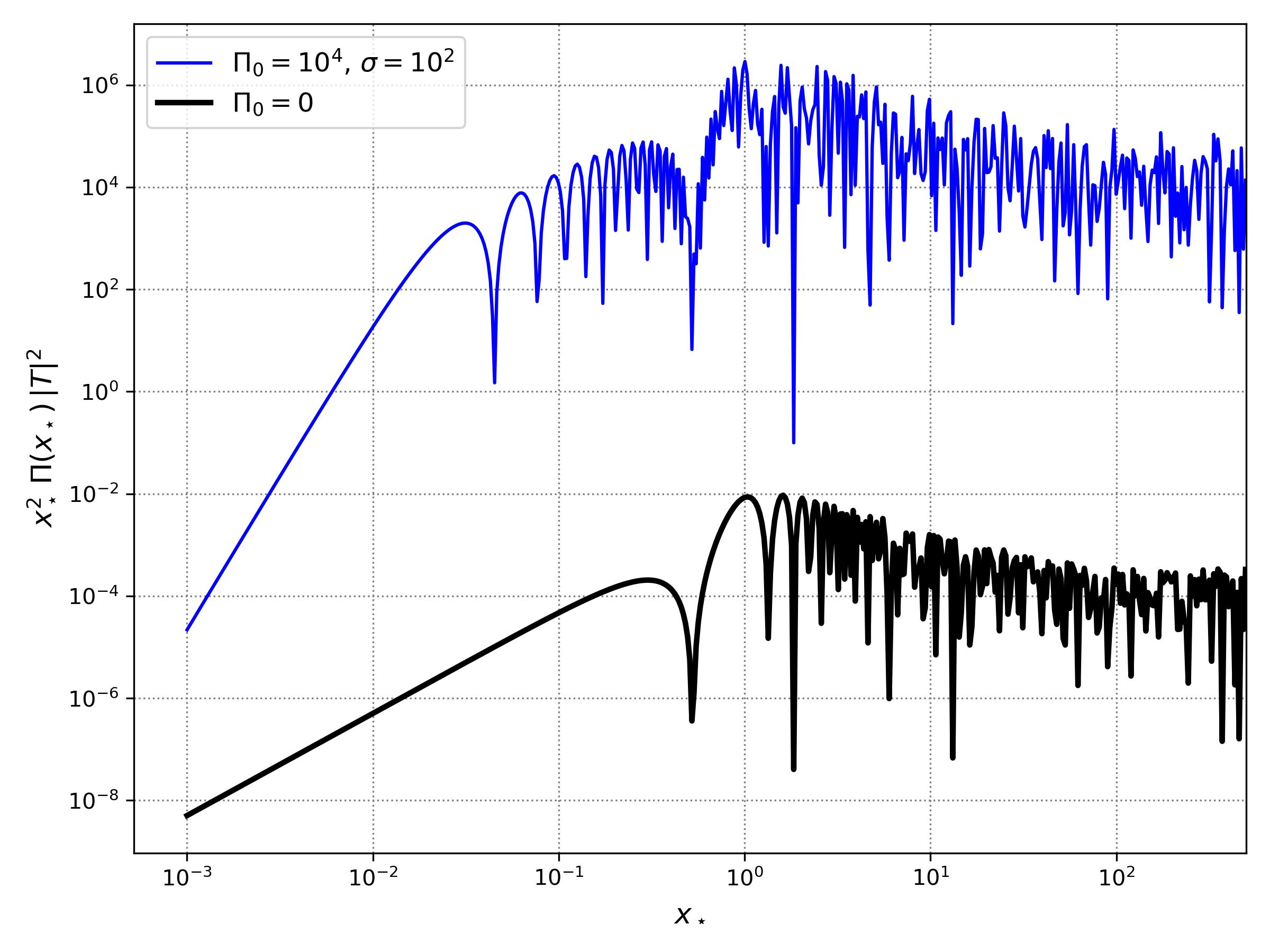}
    \caption{\small 
This series of plots shows the power spectrum ${\cal P}_{A_L}$ of the vector longitudinal mode, both during inflation and the radiation-dominated era, as a function of scale. The emphasis is on illustrating the impact in the infrared part of the spectrum of effects controlling a brief period of non-slowroll evolution during inflation. (See the main text for definitions of the parameters and further discussion.)
\textbf{Upper left panel:} The blue curve shows the profile of ${\cal P}_{A_L}$ during inflation for $\Pi_0 = 10^4$ as a function of $\kappa = k/k_1$, where $k_1$ is the characteristic scale defined in Eq.~\eqref{def_kun}. For comparison, the dashed orange line indicates the $\kappa^2$ scaling corresponding to $\Pi_0 = 0$.
\textbf{Upper right and lower panels:} The colored curves show the evolution of ${\cal P}_{A_L}$ during radiation domination for various values of the parameters $\Pi_0$ and $\sigma$, compared to the black line representing the standard profile in the absence of non-slowroll effects. Note that changes in the parameter $\sigma$ affect both the position and shape of the spectral peak. }
     \label{fig_spf1} 
\end{figure}

During the radiation domination (RD) period following inflation the large-scale longitudinal mode of the vector field, $A_L$, re-enters the horizon and begins to evolve. Its dynamics depend crucially on its comoving
momentum $k$.
  The power spectrum of the longitudinal mode at a given time $\tau\ge\tau_R$ during RD is controlled  by a transfer function ${\cal T}(k\tau)$, and is expressed as
\be
\label{ps_trd}
\mathcal{P}_{A_L}(\tau, k) = |{\cal  T}(k\tau)|^2\, \mathcal{P}_{A_L}^{(0)}(k)\,,
\ee
where $\mathcal{P}_{A_L}^{(0)}(k)$ denotes the primordial spectrum computed at the end of inflation, as described in Section~\ref{sec_inf} -- see in particular
Eq.~\eqref{NSRpowerspectrum}, as well as \cite{Marriott-Best:2025sez} for an analytical solution for the transfer functions. In the standard slowroll scenario
of \cite{Graham:2015rva}, this spectrum features a turnover at a comoving scale $k_\star \sim a_\star m$, determined by the vector mass $m$ and the scale factor $a_\star$ evaluated at the moment when the Hubble parameter satisfies $H = m$. Our goal is to investigate how a transient non-slowroll phase during inflation modifies this standard prediction --particularly through its impact on the spectrum shape, as shown in Fig.~\ref{fig_spf1}.

To analyze the evolution of the transfer function more conveniently, we introduce a set of variables following the conventions in~\cite{Ozsoy:2023gnl,Marriott-Best:2025sez}:
\bea
H_\star &=& m\hskip0.5cm,\hskip0.5cm
k_\star \,=\, a_\star m\,, \nonumber\\
\label{def_phqu}
x &=& k \tau\,\hskip0.5cm,\hskip0.5cm x_\star \,=\, {k}/{k_\star}\,, \\
y &=& \frac{x}{x_\star}\,\hskip0.5cm,\hskip0.5cm 
\sigma \,=\, \frac{a_\star H_\star}{a_1 H_I}\,.\nonumber
\eea
An analytic approximation for the transfer function ${\cal T}(k\tau) \equiv {\cal T}(x)$ in the relevant asymptotic regimes is derived in Ref.~\cite{Marriott-Best:2025sez}, to which we refer the reader for full details. That work demonstrates that ${\cal T}(x)$ can be expressed in terms of simple oscillatory functions in the two distinct regimes $x_\star < 1$ and $x_\star > 1$. We are going to  make use of this result in the following.

\smallskip

To understand the rich physical consequences of the  relevant physical scales, we recall that the comoving scale
\be
k_1 = a(\tau_1) H_I = -\frac{1}{\tau_1}
\ee
marks the onset of the non-slowroll phase during inflation. On the other hand, the scale
\be
k_\star = a_\star m
\ee
characterizes the physical momentum at which the vector mass becomes dynamically relevant during RD. We then have two distinct momentum scales to deal with. 
To proceed, it is useful to analyse further  the ratio $\sigma$ introduced
in Eq.~\eqref{def_phqu}:
\be
\sigma = \frac{k_\star}{k_1} = \frac{a_\star H_\star}{a(\tau_1) H_I} = \frac{a_\star}{a (\tau_1)} \frac{m}{H_I}\,,
\ee
which  plays a significant role in our analysis. Writing $a_R$ for the value of the scale factor at the end of inflation, we have:
\bea
\frac{a_R}{a_1} &=& e^{N_1} \,, \qquad \text{with } N_1 \text{ the number of $e$-folds between } \tau_1 \text{ and inflation end}\,, \\
\frac{a_\star}{a_R} &=& \frac{\tau_\star}{\tau_R} = \left( \frac{H_I}{H_\star} \right)^{1/2} = \left( \frac{H_I}{m} \right)^{1/2}\,.
\eea
Combining these formulas, we obtain
\be
\sigma = e^{N_1} \left( \frac{m}{H_I} \right)^{1/2}\,.
\ee
To get a sense of the possible values of $\sigma$, consider two illustrative examples:
\begin{itemize}
\item[-] For $N_1 = 30$ $e$-folds before inflation ends, $m = 10^{-19}$ eV, and $H_I = 10^{14}$ GeV, we find $e^{30} \sim 10^{13}$ and $(m/H_I)^{1/2} \sim 10^{-21}$, yielding $\sigma \sim 10^{-8}$: a negligible value.
\item[-] For $N_1 = 49$ $e$-folds  before inflation ends, $m = 10^{-19}$ eV, and $H_I = 10^{10}$ GeV, we find $e^{49} \sim 10^{21.2}$ and $(m/H_I)^{1/2} \sim 10^{-19}$, leading to $\sigma \sim 10^{2.2}$: a significantly large value.
\end{itemize}
Thus, depending on the inflationary history and the vector mass scale $m$, the parameter $\sigma$ can vary across many orders of magnitude.

\smallskip

We can now re-express the longitudinal mode spectrum of Eq.~\eqref{ps_trd}  
 in terms of the dimensionless variable $x_\star = k/k_\star$ as
\be
\label{eq_PALsq}
\mathcal{P}_{A_L}(x_\star) = \left( \frac{H_I a_\star}{2\pi} \right)^2 \left[ x_\star^2\, \Pi(\sigma x_\star)\, |{\cal T}(x_\star)|^2 \right]\,.
\ee
where ${\cal T}$ is the transfer function, and recall the definition \eqref{def_cPi} for $\Pi(x)$. Explicitly,
\be\label{def_pisxs}
 \Pi(\sigma x_\star)\,=\,1 - 4\sigma x_\star \cos(\sigma x_\star) j_1(\sigma x_\star) \Pi_0 + 4\sigma^2 x_\star^2 j_1^2(\sigma x_\star) \Pi_0^2\,,
\ee
which depends on the parameters $\sigma$ and $\Pi_0$. 
We plot the quantity in square brackets of Eq.~\eqref{eq_PALsq} in three of the panels of Fig.~\ref{fig_spf1}, noting that it is independent of the parameters $m$ and $H_I$.

\smallskip

The figures reveal a nontrivial interplay between the effects of the transfer function $|{\cal T}(x_\star)|^2$ and the deformation of the inflationary spectrum by the function $\Pi(\sigma x_\star)$. For large values of the parameter $\Pi_0$, e.g., $\Pi_0 = 10^4$ (chosen with an eye toward the applications discussed in Sections~\ref{sec_dmam} and~\ref{sec_pheno}) the longitudinal spectrum exhibits a prominent peak and a turnover at a scale dependent on the model parameters.
The amplitude of the spectrum near the peak is significantly enhanced relative to the slowroll case ($\Pi_0 = 0$), indicating that the non-slowroll epoch efficiently amplifies the production of longitudinal vector modes.   As shown in the previous section, the departure from slowroll evolution significantly impacts the infrared portion of the spectrum, enhancing its growth rate as it approaches the peak. This behavior represents a key distinction from other scenarios -- such as those involving non-standard cosmological histories after inflation, see e.g. \cite{Ahmed:2020fhc,Kolb:2020fwh}  -- which instead alter the spectral properties  \textit{after} the peak occurs.

Importantly, the location of the turnover scale is  sensitive to the value of $\sigma$. For $\sigma \sim 1$, the turnover occurs near the familiar scale $k \sim k_\star = a_\star m$, as in the standard slowroll scenario (see Fig.~\ref{fig_spf1}, upper right panel). However, in the $\sigma \ll 1$ regime, the turnover shifts to significantly smaller scales, $k \gg k_\star$, resulting in a spectrum that accumulates power at high $k$ (Fig.~\ref{fig_spf1}, lower left panel). In contrast, for $\sigma \gg 1$, the turnover remains at $k \sim k_\star$, but the small-$k$ growth of the spectrum becomes markedly steeper (Fig.~\ref{fig_spf1}, lower right panel), suggesting that the total power carried by short-wavelength modes may be significantly enhanced in this regime.

Overall, the plots in Fig.~\ref{fig_spf1} clearly illustrate the  dependence of the spectrum~\eqref{eq_PALsq} on the parameters $m$, $\Pi_0$, and $\sigma$, pointing to a rich phenomenology that we explore in the following.

\section{Longitudinal energy density and dark matter amount}
\label{sec_dmam}

Since we wish to explore the possibility that the vector longitudinal mode $A_L$
constitutes (part of) the observed dark matter
of our universe, we compute its energy density compared with the present-day
dark matter abundance. We determine how this ratio depends on the parameters
of the model, and in particular we explicitly explore
what is the role of the non-slowroll phase of evolution during inflation
for characterising the final dark matter density.

\smallskip

The energy density stored in the longitudinal vector modes  is given by \cite{Graham:2015rva}
\begin{equation}
\rho_{A_L} = \frac{m^2}{2a^2} \int d\ln k \left[ \frac{\mathcal{P}_{(\partial_{\tau}{A_L})}(k)}{k^2 + a^2 m^2} + \mathcal{P}_{A_L}(k) \right]\,,
\label{density}
\end{equation}
where the first term corresponds to the kinetic energy and the second to the potential energy. Both terms evolve in time via the transfer function. They share a common overall dependence on the primordial power spectrum $\mathcal{P}_{A_L}^{(0)}$
we analyzed in section \ref{sec_inf}. Plugging the explicit form of $\mathcal{P}_{A_L}^{(0)}$ obtained from a transient non-slowroll phase during inflation—encoded through the time-dependence of the vector mass—we obtain an
integral involving transfer functions
\begin{align}
\rho_{A_L} &= \frac{1}{2a^2} \frac{a_\star}{a} \left( \frac{k_\star H_I}{2\pi} \right)^2 
\left[
\int d\ln k \left( \frac{k^2}{k_\star^2(k^2 + a^2 m^2)} \frac{a}{a_\star} |\partial_\tau \mathcal{T}|^2 + \frac{k^2}{k_\star^2} \frac{a}{a_\star} |\mathcal{T}|^2 \right)
\Pi(k \tau_1)
\right]
\label{integral}
\end{align}
where the explicit oscillatory structure in the last factor $\Pi(k \tau_1)$ of the integrand arises from   the non-slowroll transition at time $\tau_1$, and correspondingly introduces a physical scale $k_1 = a(\tau_1) H_I$: see the definition in Eq.~\eqref{def_cPi}.

\subsubsection*{Computing the integral}

We call ${\cal I}_\rho$ the dimensionless integral in the square parenthesis of Eq.~\eqref{integral}. By  making
use of the dimensionless variables of Eq.~\eqref{def_phqu}, 
  and evaluating it  at late times $y_{\rm end}\, =\, a(\tau_{\rm end})m/k_\star \gg 1$,
  we write
\begin{align}
{\cal I}_\rho(y_{\rm end}) &= y_{\rm end} \int d\ln x_\star \left( \frac{x_\star^2}{x_\star^2 + y^2_{\rm end}} |\partial_y \mathcal{T}|^2 + x_\star^2 |\mathcal{T}|^2 \right) \Pi(\sigma x_\star)\,.
\end{align}
We choose a large value of $y_{\rm end}$ in order to evaluate quantities
deep in the radiation dominated era, and to average over rapid oscillations. 
Recalling that $\sigma = k_\star \tau_1$,
 the modulation function inside the integral is given by Eq.\eqref{def_pisxs}. 
 Following the approach of~\cite{Marriott-Best:2025sez}, we split the integral into two regions, $x_\star < 1$ and $x_\star > 1$.
They correspond respectively to modes outside and inside the Hubble radius during radiation domination:
\begin{align}
\mathcal{I}_\rho(y_{\rm end}) &= y_{\rm end} \int_0^1 \frac{dx_\star}{x_\star} \left( \frac{x_\star^2}{x_\star^2 + y_{\rm end}^2} |\partial_y T^{(A)}_{\rm late}|^2 + x_\star^2 |T^{(A)}_{\rm late}|^2 \right) \Pi(\sigma x_\star) \nonumber \\
&\quad + y_{\rm end} \int_1^{y_{\rm end}} \frac{dx_\star}{x_\star} \left( \frac{x_\star^2}{x_\star^2 + y_{\rm end}^2} |\partial_y T^{(B)}_{\rm late}|^2 + x_\star^2 |T^{(B)}_{\rm late}|^2 \right) \Pi(\sigma x_\star)\,.
\label{int2pie}
\end{align}
Each line of Eq.~\eqref{int2pie} depends on the different solutions for the analytic transfer
functions which apply  in different regimes of $y$ (see Section 2.3 of the arXiv version of
\cite{Marriott-Best:2025sez}):
\begin{enumerate}
\item
The first line of Eq~\eqref{int2pie}, call it $\mathcal{I}^{(1)}$, can be easily evaluated  for large values of the argument $y_{\rm end}$. We find
\begin{equation}
\mathcal{I}^{(1)} = 0.475883 + s_1(\sigma)\, \Pi_0 + s_2(\sigma)\, \Pi_0^2\,.
\label{int_firlin}
\end{equation}
The expressions for  $s_1(\sigma)$ and $s_2(\sigma)$
are long and relegated to Appendix \ref{app_C}. Sufficient to say 
that
they 
 vanish as $\sigma \to 0$ and asymptote to the same constant $\sim0.95$ for large $\sigma$: see the left panel
 of Fig \ref{fig_cofla2}.

\item 
 For dealing with the  contribution in the second
 line of  Eq~\eqref{int2pie}, call it $\mathcal{I}^{(2)}$, we proceed as in \cite{Marriott-Best:2025sez}.  Given that the quantity $y_{\rm end}$ occurs both as an integration limit and within the integrand, it is convenient to first perform a rescaling of the integration variable by defining $z \equiv \frac{x_*}{y_{\rm end}}$. In terms of this new variable, the lower integration limit $\frac{1}{y_{\rm end}}$ becomes small enough to be approximated by zero. Moreover, the resulting integrand contains terms that oscillate rapidly within the interval $0 \leq z \leq 1$, and these oscillations are suppressed by inverse powers of $y_{\rm end}$. Since these contributions average out to nearly zero, they can be safely neglected. The result can be handled exactly, 
and
 we obtain the following
 expression 
\begin{equation}
\mathcal{I}^{(2)} = 1.025 \left[ 1 + 2 \Pi_0 + \left(2 + \frac{0.685}{\sigma^2} \right) \Pi_0^2 \right]\,.
\end{equation}
\end{enumerate}

\subsubsection*{Collecting the results}

Let us collect the results obtained so far. 
 The total integral  \eqref{int2pie} is the sum of the previous
two pieces: $\mathcal{I}_\rho\, = \,\mathcal{I}^{(1)}+\mathcal{I}^{(2)}$, and depends
both on $\sigma$ and $\Pi_0$. Interestingly,  $\mathcal{I}_\rho$ 
acquires a particularly simple expression for
 large $\sigma$
\begin{equation}
\label{exp_is}
\mathcal{I}_\rho\,\simeq\, \frac{3}{2} \left(1 + 2 \Pi_0 + 2 \Pi_0^2 \right)\,
\hskip0.6cm {\text{for large $\sigma$.}}
\end{equation}
\begin{figure}[t]
    \centering
          \includegraphics[width=0.52\linewidth]{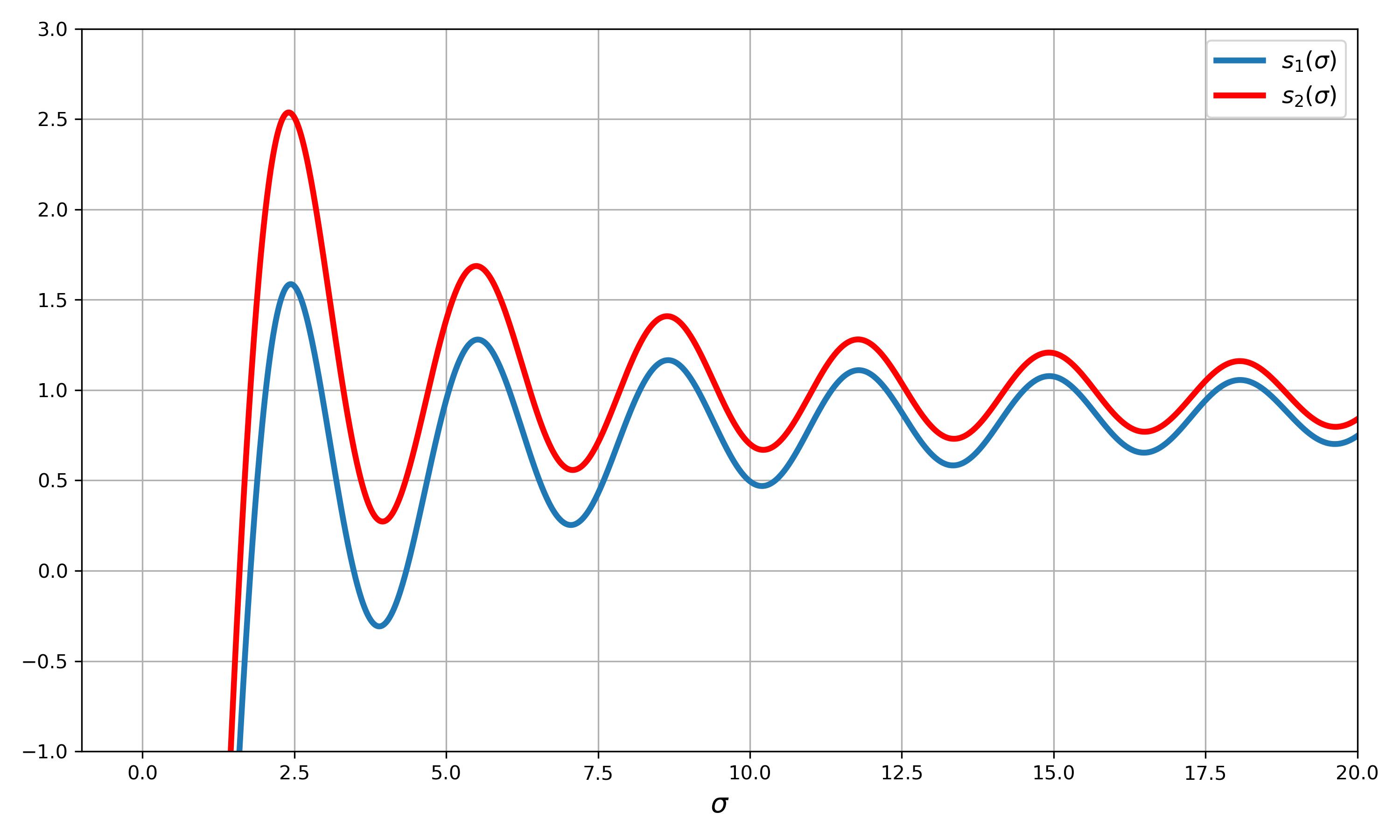}
                \includegraphics[width=0.43\linewidth]{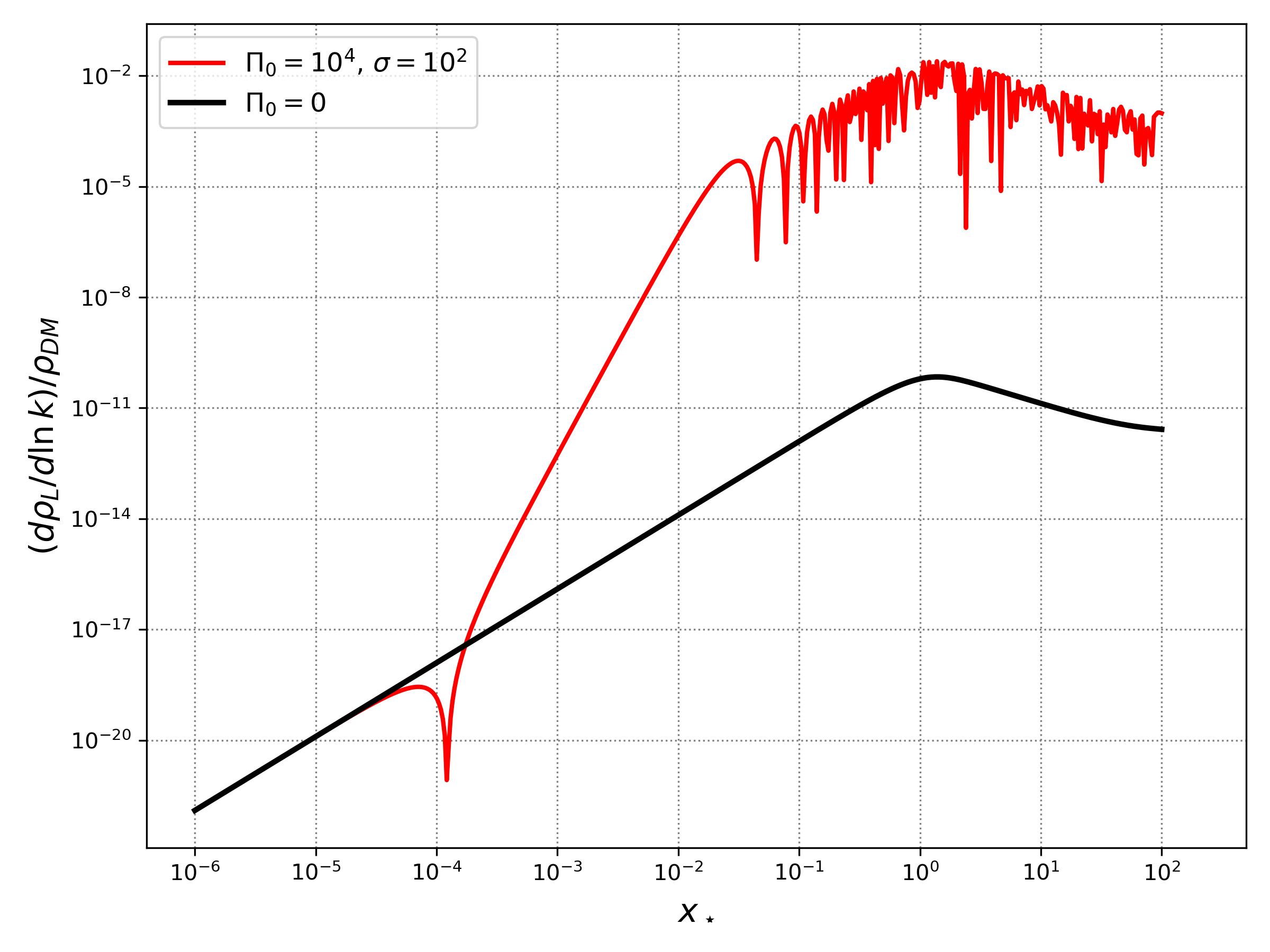}
    \caption{\small
\textbf{Left panel:} Plot of the functions $s_{1,2}(\sigma)$ defined in Eq.~\eqref{int_firlin}. Although these functions exhibit oscillations, they asymptotically approach a common value for large $\sigma$. \textbf{Right panel:} Plot of the differential energy density \eqref{eq_diffend} as a function of momentum scale, for a representative choice of vector mass $m = 1.5 \times 10^{-19}\, \text{eV}$ and inflationary Hubble scale $H_I = 10^{14}\, \text{GeV}$. The figure highlights how the parameters $\sigma$ and $\Pi_0$, which control the non-slowroll dynamics during inflation, significantly enhance the amplitude of the peak in the spectrum.  }
    \label{fig_cofla2} 
\end{figure}
The resulting energy density in longitudinal vector modes, which redshifts as matter, is given by
\begin{equation}
\rho_{A_L} = \frac{1}{2} \frac{a_\star^3}{a^3} \left( \frac{m H_I}{2\pi} \right)^2 \mathcal{I}_\rho\,.
\end{equation}
Evaluating this formula at matter-radiation equality and comparing with the total dark matter density \cite{Ade:2015xua} 
\begin{equation}
\rho_{\rm DM} = \frac{3}{2} H_{\rm eq}^2 M_{\rm Pl}^2\,, \quad H_{\rm eq} = 2.8 \times 10^{-28}\,{\rm eV}\,,
\end{equation}
we obtain, in a large $\sigma$ regime, the ratio
\begin{equation}
\frac{\rho_{A_L}}{\rho_{\rm DM}} \,=\, \frac{m^{1/2}}{3 H_{\rm eq}^{1/2}} \left( \frac{H_I}{2\pi M_{\rm Pl}} \right)^2 \mathcal{I}_\rho \,
=\, \left(1 + 2\Pi_0 + 2\Pi_0^2 \right) \left( \frac{m}{0.6 \times 10^{-6}\,{\rm eV}} \right)^{1/2} \left( \frac{H_I}{10^{14}\,{\rm GeV}} \right)^2\,.
\label{eq_ratioen}
\end{equation}
Hence, the result depends in a rich way on the available parameters. As explicit examples,
we can consider a regime where the longitudinal vector plays the the role of ultra light
dark matter candidate. Choosing
 $m = 1.5 \times 10^{-21}\,{\rm eV}$ and $H_I = 10^{14}\,{\rm GeV}$, a value $\Pi_0^2 = 10^7$ yields the correct dark matter abundance. As an additional 
case, choosing instead $H_I = 5\times 10^{13}\,{\rm GeV}$ (keeping the same values
 for the other parameters),  we obtain a scenario where the longitudinal vector is an ultralight
dark matter candidate constituting the 25\% of  the observed dark matter in the universe. 
As we shall see in the next Section, besides known interesting features as ultralight dark matter candidate (see the recent reviews \cite{Ferreira:2020fam,Eberhardt:2025caq}), the very small mass  range considered above has  ramifications for the physics of gravitational waves.
Formula \eqref{eq_ratioen} is similar in spirit to analog equations recently obtained in a phenomenological way in \cite{Marriott-Best:2025sez}: in fact, by means of  our construction based on non-slowroll dynamics during inflation, we are finding  an explicit realization of the ideas proposed in \cite{Marriott-Best:2025sez}.
It is important to emphasize that the mechanism of ultralight DM production we propose differs fundamentally from other scenarios, such as those based on the misalignment mechanism (see \cite{Ferreira:2020fam,Eberhardt:2025caq} for reviews). In contrast to these approaches, our framework does not rely on the presence of background fields oscillating around the minimum of their potential.

\smallskip

To conclude this section, we discuss an informative quantity that provides insight into how the energy density of the longitudinal mode is distributed across momentum scales. Specifically, we consider the differential contribution to the energy density, normalized by the total dark matter density:
\begin{equation}
\label{eq_diffend}
\frac{1}{\rho_{\rm DM}} \frac{d\rho_{A_L}}{d\ln k}\,.
\end{equation}
This quantity represents the contribution per logarithmic interval in momentum to the energy density of the longitudinal vector mode. When integrated over all momentum scales, it yields the total energy density ratio given in Eq.~\eqref{eq_ratioen}. However, before integration, it offers an instructive view of how different momentum modes contribute to the overall energy density, thereby serving as a diagnostic for scale-dependent features in the production mechanism.
In the right panel of Fig.~\ref{fig_cofla2}  we plot this distribution. One can clearly see that the brief non-slowroll phase during inflation has a significant effect on the infrared part of    the spectrum,  in the region where it  is growing. The modification of the profile in this growing region reflects the enhanced amplification of certain modes due to the departure from slowroll evolution. 

\section{Phenomenological implications}
\label{sec_pheno}

The scenario for the inflationary production of longitudinal vector dark matter developed in the previous sections exhibits several appealing features. It generates a spectrum of longitudinal modes that is strongly suppressed on large scales -- thus evading stringent constraints from cosmic microwave background observations -- and grows toward smaller scales, reaching a peak whose position and amplitude depend on the model parameters: the vector mass $m$, and the parameters $\sigma$ and $\Pi_0$ characterizing the brief non-slowroll phase during inflation. The infrared part of the spectrum is especially sensitive to the inflationary non-slowroll phase. 

Remarkably, this framework allows for the efficient production of ultralight vector dark matter from inflation, with masses comparable or smaller than $m = 10^{-19}$ eV, while still yielding an energy density sufficient to constitute a substantial fraction of the total dark matter abundance. Ultralight dark matter has distinctive phenomenological implications, as its wave-like nature can influence the formation and evolution of cosmic structures at small scales, possibly addressing some of the problems of more standard dark matter scenarios. Among others, recent
works \cite{Dalal:2022rmp,Zimmermann:2024xvd} set   bounds on the mass
of generic ultralight dark matter candidates, 
if they constitute the totality of the dark matter. 
For a comprehensive discussion, we refer the reader to the recent reviews~\cite{Ferreira:2020fam,Eberhardt:2025caq}. In the form of dark
photon, a light vector dark matter can also have interesting consequences through possible kinetic mixing with Standard Model particles, see \cite{Fabbrichesi:2020wbt} for a review. Hence our mechanism offers
a new tool to produce ultralight dark matter from the early Universe.
In our context, as we are going to discuss, the small-mass regime has also additional implications for gravitational wave physics. 

\smallskip

  We examine the production of a spectrum of induced gravitational waves generated at second order in perturbation theory from the amplified, peaked spectrum of longitudinal vector modes at small scales (see the right panel of Fig.~\ref{fig_cofla2}).
The enhancement of the longitudinal vector spectrum due to the non-slowroll phase is expected to significantly boost the generation of gravitational waves -- potentially to an amplitude detectable by future experiments. The study of scalar-induced gravitational waves sourced by adiabatic curvature perturbations has a long  history; see, for instance, \cite{Matarrese:1992rp,Ananda:2006af,Baumann:2007zm,Saito:2009jt,Bugaev:2009zh,Espinosa:2018eve,Kohri:2018awv}, and the comprehensive review \cite{Domenech:2021ztg}. In contrast, comparatively less attention has been given to gravitational waves sourced by non-adiabatic perturbations; see e.g.\cite{Domenech:2021and,Passaglia:2021jla,Domenech:2023jve,Ozsoy:2023gnl}.
The formalism required to study the case of non-adiabatic longitudinal vector modes -- accounting for the non-standard kinetic structure of the action governing our system -- has recently been developed in \cite{Marriott-Best:2025sez}. In what follows, we make use of those results and refer the reader to that work for technical details.
\,
The gravitational wave energy density as a function of momentum scale $k$ is given by
\begin{equation}
\label{intOGW2}
\Omega_{\rm GW}(k)
=
\frac{m^4\,k^4}{12\,k_\star^4}
\int_0^{\infty} dt \int_0^1 ds\,
\left[
\frac{t(2+t)(s^2 - 1)}{(1 - s + t)(1 + s + t)}
\right]^2
\,\bar{\cal I}_{cs}^2(k, u, v)\,
{\cal P}_\varphi^{(0)}(uk)\,{\cal P}_\varphi^{(0)}(vk)\,,
\end{equation}
where the variables $u$ and $v$ are defined by $u = (t + s + 1)/2$ and $v = (t - s + 1)/2$, and $k_\star = a_\star m$ as introduced in Eq.~\eqref{def_phqu}. In what follows, we often use the rescaled variables defined there for convenience.

The kernel $\bar{\cal I}_{cs}^2(k, u, v)$ corresponds to a time average over fast oscillations -- a standard approach in the computation of induced gravitational wave spectra  -- and takes the form:
\begin{equation}
\label{timein5}
\bar{\cal I}_{cs}^2(x_\star, u, v)
=
\left| 
\int_0^{y/k_\star} y_1\,dy_1\,\cos(x_\star y_1)\,\beta(y_1,x_\star,u,v)
\right|^2
+
\left|
\int_0^{y/k_\star} y_1\,dy_1\,\sin(x_\star y_1)\,\beta(y_1,x_\star,u,v)
\right|^2.
\end{equation}
The function $\beta$ contains the transfer functions for the longitudinal vector modes during radiation domination:
\begin{equation}
\label{defff}
\beta(\tau, {\bf k}, {\bf \tilde{k}}) \equiv 
{\cal T}(\tilde{k} \tau)\,{\cal T}(|{\bf k} - {\bf \tilde{k}}| \tau)
-
\frac{a^2 M^2 \tilde{k} |{\bf k} - {\bf \tilde{k}}|\, {\cal T}'(\tilde{k} \tau)\, {\cal T}'(|{\bf k} - {\bf \tilde{k}}| \tau)}{
\left(\tilde{k}^2 + a^2 m^2\right)\left(|{\bf k} - {\bf \tilde{k}}|^2 + a^2 m^2\right)}.
\end{equation}
The size of momenta ${\bf k}$ and ${\bf \tilde{k}}$ relate to the variables $u$ and $v$ via $v = \tilde{k}/k$ and $u = |{\bf k} - {\bf \tilde{k}}|/k$.

\smallskip

\begin{figure}[t!]
    \centering
    \includegraphics[width=0.5\linewidth]{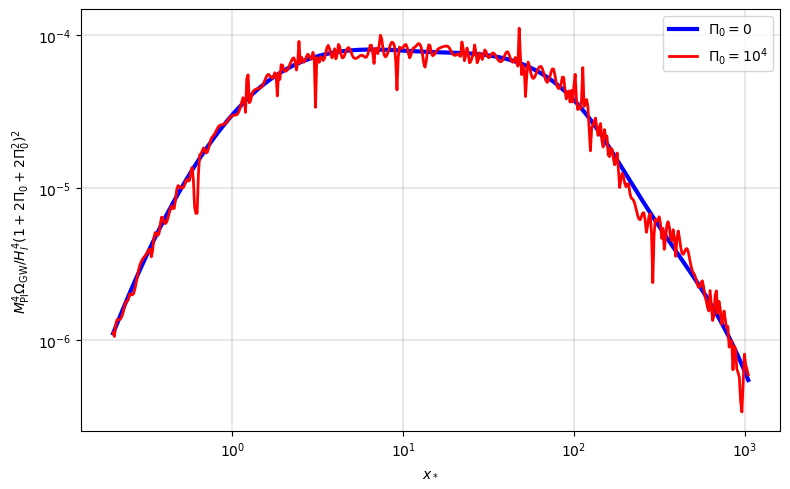}
    \caption{\small Spectrum of induced stochastic gravitational wave background numerically obtained from Eq.~\eqref{intOGW2}, and shown for two values of the parameter $\Pi_0$, with $\sigma = 10^2$ fixed.}
    \label{fig_OGWv1}
\end{figure}

The function ${\cal P}_\varphi^{(0)}$ appearing in the integrand of Eq.~\eqref{intOGW2} is the same as introduced in Section~\ref{sec_genev}, and is directly related to the longitudinal vector spectrum.  (We refer the reader to that section for recalling the variables we
are going to use next.)
It is expressed in terms of the rescaled variable $x_\star = k/k_\star$ as:
\begin{equation}
\label{spec_ls}
{\cal P}_\varphi^{(0)}(x_\star) =
\frac{H_I^2}{4\pi^2 m^2}
\left[
1 - 4\sigma x_\star \Pi_0 \cos(\sigma x_\star)\, j_1(\sigma x_\star)
+ 4\sigma^2 x_\star^2 \Pi_0^2 j_1^2(\sigma x_\star)
\right],
\end{equation}
where $\Pi_0$ and $\sigma$ are the parameters introduced in Section~\ref{sec_evo}, controlling respectively the strength and location of the features associated with the non-slowroll phase during inflation.
The oscillatory structure of Eq.~\eqref{spec_ls} complicates the numerical evaluation of the integral in Eq.~\eqref{intOGW2}. Using the methods of \cite{LISACosmologyWorkingGroup:2025vdz}, we performed a numerical integration of this expression in the regime of large $\sigma$ and $\Pi_0$. The result is shown in Fig.~\ref{fig_OGWv1}, plotted as a function of $f/f_\star$ with $k = 2\pi f$. The spectrum exhibits small-scale oscillatory features, which we interpret as numerical artifacts and therefore disregard in our analysis.

\smallskip

In the large-$\sigma$ regime, we find that the numerical results for the gravitational-wave density are well fitted by the following analytical expression for $\Omega_{\rm GW}$ as a function of frequency:
\begin{equation}
\label{eq_OGWfit}
\Omega_{\rm GW}(x_\star)
\,=\,1.2 \times 10^{-24}\,\left( \frac12 + \Pi_0 + \Pi_0^2\right)^2
\left(\frac{H_I}{10^{14}\,{\rm GeV}} \right)^4\,
\,\frac{x_\star^{2.6}}{(1 + x_\star^2)^{1.325}}
\left(1 + \frac{x_\star^3}{7.29 \times 10^5} \right)^{-0.65}\,,
\end{equation}
where again $x_\star = f/f_\star$. This fitting formula accurately reproduces the shape of the gravitational wave spectrum shown in Fig.~\ref{fig_OGWv1} across the relevant frequency range, averaging out its small oscillations. It is important to notice that  $\Omega_{\rm GW}$, as function of frequency, exhibits a plateau ranging approximately between $5\le f/f_\star\le 50$ \footnote{Besides the analysis carried
on so far in the large $\sigma$ limit, it would also be interesting to study in detail
the behaviour of the system for intermediate or small values of $\sigma$: the numerics
involved for dealing with integral \eqref{intOGW2} becomes much harder to deal with, and we prefer to postpone such analysis to future works.}.

\smallskip

\begin{figure}[t!]
    \centering
    \includegraphics[width=0.6\linewidth]{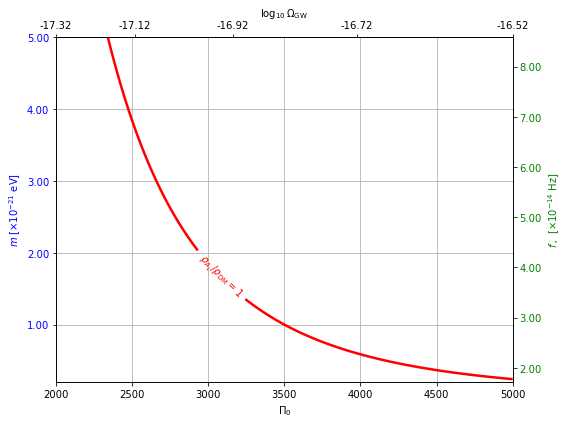}
    \caption{\small Plot summarizing Eqs.~\eqref{eq_fsta}, \eqref{eq_ratioen2} and \eqref{eq_OGWfit2} for a scenario in which longitudinal modes account for all of the dark matter. By varying the parameter $\Pi_0$ over
    a certain interval, this condition can be satisfied through corresponding changes in the vector mass. As a byproduct, an induced stochastic gravitational-wave background is generated, with amplification at very low frequencies. We choose 
${H_I}={10^{14}\,{\rm GeV}}$.}
    \label{fig_contour}
\end{figure}
Let us further discuss some of the properties of the resulting gravitational
wave spectrum, building on \cite{Marriott-Best:2025sez}. We re-collect  previous formulas
with the aim to determine the allowed amplitude and position of the plateau of  $\Omega_{\rm GW}$ in Fig.~\ref{fig_OGWv1}. The plateau position lies at $f\sim f_\star$, where $f_\star = k_\star/(2\pi)$: expressed in Hertz, this quantity reads
\be
\label{eq_fsta}
f_\star
\,=\,4.7\times 10^{-14}\,\left( \frac{m}{1.5\times 10^{-21}\,{\rm eV}}\right)^{1/2}
\,{\rm Hz}
\ee
As we learned in Sec.~\ref{sec_evo}, the relative abundance of  longitudinal vector
modes relative to present-day dark matter density reads, in a large $\sigma$ regime,
\be
\frac{\rho_{A_L}}{\rho_{\rm DM}}  \,
=\, 10^{-7}\left(\frac12 + \Pi_0 + \Pi_0^2 \right) \left( \frac{m}{1.5 \times 10^{-21}\,{\rm eV}} \right)^{1/2} \left( \frac{H_I}{10^{14}\,{\rm GeV}} \right)^2\,.
\label{eq_ratioen2}
\ee
The maximal value of $\Omega_{\rm GW}$ at the position of the plateau is approximately
given by the overall coefficient of Eq.~\eqref{eq_OGWfit},
\begin{equation}
\label{eq_OGWfit2}
\Omega_{\rm GW}^{\rm max}
\,=\,1.2 \times 10^{-24}\,\left( \frac12 + \Pi_0 + \Pi_0^2\right)^2
\left(\frac{H_I}{10^{14}\,{\rm GeV}} \right)^4
\,.
\end{equation}
For definiteness, let us fix the Hubble scale at ${H_I} = 10^{14}\,{\rm GeV}$. 
From Eq.~\eqref{eq_OGWfit2}, we notice that the amplitude of the gravitational wave spectrum increases with the parameter $\Pi_0$. 
A large value of $\Pi_0$ is permitted by Eq.~\eqref{eq_ratioen2} in the regime of small vector mass. 
This makes the ultralight dark matter regime particularly relevant for our purposes. 
For instance, a mass $m = 1.5 \times 10^{-21}\,{\rm eV}$ requires $\Pi_0^2 \simeq 10^7$ for the longitudinal vector mode to account for (most of) the observed dark matter, as dictated by Eq.~\eqref{eq_ratioen2}. 
The corresponding maximal gravitational wave energy density, $\Omega_{\rm GW}^{\rm max}$, reaches a relatively large amplitude of order $10^{-10}$. 
However, according to Eq.~\eqref{eq_fsta}, the associated plateau occurs at a very low frequency, of order $f_\star \simeq 10^{-14}\,{\rm Hz}$, which lies in an intermediate range between the scales accessible to cosmic microwave background polarization experiments, and those probed by pulsar timing arrays. 
At present, there are no experiments  able to detect gravitational waves at these frequencies. 
Nevertheless, future experiments aiming to measure redshift distortions in the microwave background could potentially be sensitive to tensor modes in this range. 
One such proposal is SuperPIXIE \cite{Kite:2020uix}, an evolution of the PIXIE experiment \cite{Kogut:2011xw,Kogut:2019vqh,Chluba:2019nxa}. Also, this frequency range
might be probed in terms of gravitational wave-induced distortions
of galaxy shapes, see e.g. \cite{Schmidt:2013gwa,Biagetti:2020lpx}.

\smallskip

To conclude this section, we point out additional  compelling aspect of our scenario. With suitable extensions, it can provide a unified framework for producing both ultralight vector dark matter and primordial black holes. The common ingredient is the short-lived violation of slowroll conditions during inflation. While our focus here has been on how this non-slowroll phase induces a rapid change in the effective vector mass -- leading to enhanced longitudinal mode production -- the same mechanism also amplifies adiabatic curvature perturbations. These amplified perturbations can, in turn, lead to the formation of primordial black holes during the radiation-dominated era.
This opens the possibility for a mixed dark matter scenario, composed of both ultralight vector modes and primordial black holes, with the mass spectrum of the latter determined by the properties of the non-slowroll phase. Such a framework could help evade stringent observational constraints that apply when each dark matter candidate is considered independently.  Moreover, it can lead to a gravitational wave background with a bimodal shape, with two peaks caused by the amplified isocurvature and adiabatic fluctuations 
at different frequencies.
We consider this an intriguing direction for further research, which we intend to investigate in a forthcoming publication.

\section{Conclusions}

 In this work, we have presented a generalization of the scenario proposed in~\cite{Graham:2015rva}, constructing a model of longitudinal vector dark matter produced during inflation that incorporates a brief phase of non-slowroll evolution. If the vector mass depends on the inflaton velocity, it can undergo substantial variations during this non-slowroll phase, due to rapid changes in the time derivatives of the inflaton field. We have shown that such a phase can significantly modify the infrared behavior of the longitudinal mode spectrum, enhancing its growth to a $k^6$ scaling—compared to the $k^2$ rise characteristic of the original slow-roll setup.

\smallskip

This modification has noteworthy consequences, as it permits the vector field mass to be as small as around \( m \lesssim 10^{-20}~\mathrm{eV} \) or even lower, while still accounting for a fraction of the observed dark matter abundance in the form of longitudinal vector bosons. It is worth stressing that the mechanism of ultralight DM production we propose differs fundamentally from other scenarios, such as those based on the misalignment mechanism (see \cite{Ferreira:2020fam,Eberhardt:2025caq} for reviews). In contrast to these approaches, our framework does not rely on the presence of background fields oscillating around the minimum of their potential. Furthermore, building on~\cite{Marriott-Best:2025sez}, we demonstrated that this scenario naturally leads to a stochastic background of gravitational waves with relatively large amplitude, generated at second order in perturbation theory, sourced by the enhanced small-scale vector spectrum. The resulting signal  is peaked at extremely low frequencies, \( f \sim 10^{-15} \)--\(10^{-13}~\mathrm{Hz} \), making it
a distinctive prediction of our ultralight dark matter model, which
can be probed even if dark matter has no interactions with the Standard Model
besides gravity.  
%
%
%
An intriguing feature of this framework is its potential ability to simultaneously generate both longitudinal vector dark matter and primordial black holes, offering a unified mechanism for producing a two-component dark matter model from a common inflationary source.

\smallskip

Several directions remain open for future investigation. It would be valuable to identify explicit particle physics realizations of this ultralight vector dark matter scenario, potentially within extensions of the Standard Model that accommodate very light vector bosons. It is also important to further explore observational strategies for detecting ultra-low-frequency gravitational waves as predicted by this mechanism, for example through galaxy shape correlations and intrinsic alignments. We plan to report on these developments in future work.

\subsection*{Acknowledgments}
It is a pleasure to thank  Alisha Marriott-Best, Ogan \"Ozsoy,  Marco Peloso,  Davide Racco and Ivonne Zavala
 for useful discussions and feedback.
We are partially funded by the STFC grants ST/T000813/1 and ST/X000648/1. 
For the purpose of open access, the authors have applied a Creative Commons Attribution licence to any Author Accepted Manuscript version arising. Research Data Access Statement: No new data were generated for this manuscript.

\begin{appendix}
\section{An explicit inflationary setup}
\label{app_mod}

We present a setup for inflation which can generate vector field mass that rapidly varies for a short amount of time. Our setup is motivated by Starobinsky model
\cite{Starobinsky:1992ts} 
with singularities in
the inflaton 
potential.
The complete action for the system during inflation  is
\be
\label{eq_dia2}
S = \int d^4 x\, \sqrt{-g} \left[
\frac{R}{2} - \frac{1}{4} F_{\mu\nu} F^{\mu\nu} - \frac{M^2(\dot \phi)}{2} A_\mu A^\mu
-\frac12 \partial_\mu \phi \partial^\mu \phi -V(\phi)\right]\,.
\ee
For convenience, in this appendix we  work with physical time, related to
conformal time by $d \tau = d t/a(t)$. The potential for the inflaton field $\phi$
is
parametrized as
\bea
\label{ans_omeV}
V(\phi) = 
\begin{cases}
V_0+{b_1}\,{H_I^3}\, (\phi-\phi_0)& \text{for } \phi > \phi_0\,, \\
V_0+{b_2}\,{H_I^3}\, (\phi-\phi_0)& \text{for } \phi > \phi_0\,, 
\end{cases}
\eea
with $H_I$ the (nearly constant) Hubble parameter during inflation,
$V_0$ a constant scale, and $b_{1,2}$ two positive 
parameters satisfying the hierarchy $b_1\gg b_2$. This potential
is continuous but its first derivative is discontinuous at $\phi=\phi_0$. 
The  vector field mass in action \eqref{eq_dia2} is chosen as
\be
M^2(t)\,=\,m^2\,\frac{\dot \phi^2(t)}{H_I^4}
\,,\ee
with $m$ a constant mass parameter.
It is straightforward to show (see, e.g., \cite{Leach:2001zf}) that, during quasi-de Sitter expansion, the evolution equation of the inflaton scalar field leads to the following relation for its velocity, where $t_0$ denotes the time at which $\phi(t_0) = \phi_0$:
\bea
\label{eq_velo}
3 H_I \dot \phi(t) \simeq 
\begin{cases}
-{b_1}\,{H_I^3} & \text{for } t < t_0\,, \\
-{b_2}\,{H_I^3}- (b_1-b_2)\,{H_I^3}\,e^{-3 H_I (t-t_0)} & \text{for } t \ge t_0\,.
\end{cases}
\eea
Hence the scalar field velocity decreases exponentially
fast for a short period of time starting at $t_0$. Correspondingly,
also the vector mass quickly decreases its value
at around that epoch, passing from $M^2 = m^2\,b_1^2/9$ 
for  $t < t_0$
to  $M^2 = m^2\,b_2^2/9$ at  $t > t_0$.
Hence Starobinsky model \cite{Starobinsky:1992ts}
offers a possible explicit framework for developing
the mechanism studied in this work.

\section{The values of $s_1$ and $s_2$ in Eq~\eqref{int_firlin}}
\label{app_C}
In this Appendix we report the expressions for the quantities $s_1$ and $s_2$ appearing in Eq~\eqref{int_firlin} as a function of $\sigma$. The symbol $ \text{Ci}$ indicates
the cosine integral, and $\gamma_E$ the Euler-Mascheroni constant.
\bea
s_1(\sigma)&=&
\frac{1}{16 \sigma^2 (-1 + \sigma^2)^2}
\Big[
4 \Big( 
    -3 + \sigma^2 \left( 
        4 + 5 \gamma_E (-1 + \sigma^2)^2 - \sigma^2 (2 + \sigma^2) 
    \right)
    + 3 \cos(2\sigma) 
    \nonumber
    \\
&&    + \sigma \Big[ 
        \sigma \left( 
            -6 + \cos(2) + \sigma^2 (3 + \cos(2)) 
        \right) \cos(2\sigma) 
       \nonumber
    \\
&&     + (-1 + \sigma^2)^2 \left( 
            -6 \, \text{ArcCoth}(\sigma) + 3 \sigma \cos(2) - 5 \sigma \, \text{Ci}(2) 
        \right)
    \Big] 
\Big)
 \nonumber
    \\
&& 
+ 2 \sigma \Big[ 
    -4 \sigma (-1 + \sigma^2) \cos(2\sigma) \sin(2)  \nonumber
    \\
&& 
    + (-1 + \sigma^2)^2 \Big( 
        (-6 + 5\sigma) \, \text{Ci}(2(1 - \sigma)) + 6 \, \text{Ci}(2(1 + \sigma))
  \nonumber
    \\
&&    + \sigma \left( 
            -10 \, \text{Ci}(2\sigma) + 5 \, \text{Ci}(2(1 + \sigma)) 
            + \log(1024) + 10 \log(\sigma) - 5 \log(-1 + \sigma^2) + 4 \sin(2) 
        \right)
    \Big] \nonumber
    \\
&& 
    -2 \left[ 
        (-1 + \sigma^2)(7 - 5 \cos(2) + \sigma^2(-7 + 3 \cos(2))) 
        + 2(1 - 3\sigma^2 + \sigma^4) \sin(2) 
    \right] \sin(2\sigma) 
\Big)\Big]\,,
\\
s_2(\sigma)&=&
\frac{1}{8 \sigma^2 (1 - \sigma^2)^2}
\Big[
-15 + 30 \sigma^2 - 17 \sigma^4 - 2 \sigma^6 
+ 2 \gamma_E (1 - \sigma^2)^2 (4 + 5 \sigma^2) 
+ (1 - \sigma^2)^2 (5 + 6 \sigma^2) \cos(2) 
\nonumber
\\
&&
+ \left[15 (1 - \sigma^2)^2 + (-5 + 8 \sigma^2 + \sigma^4) \cos(2) \right] \cos(2\sigma) 
+ 12\, \mathrm{Ci}(2) 
\nonumber
\\
&&
- 2 \sigma^2 (17 - 16 \sigma^2 + 5 \sigma^4) \mathrm{Ci}(2) 
+ (1 - \sigma^2)^2 (-6 + 5 \sigma^2) \mathrm{Ci}(2(1 - \sigma)) 
- 2 (1 - \sigma^2)^2 (4 + 5 \sigma^2) \mathrm{Ci}(2 \sigma) 
\nonumber
\\
&&
- 6\, \mathrm{Ci}(2(1 + \sigma)) 
+ \log(256) + 8 \log(\sigma) + 6 \log(1 - \sigma^2) 
- 2 \sin(2) 
- 2 (1 - \sigma^4) \cos(2\sigma) \sin(2) 
\nonumber
\\
&&
+ \sigma^2 \Big[
    (17 - 16 \sigma^2 + 5 \sigma^4) \mathrm{Ci}(2(1 + \sigma)) 
    + 2 (-3 - 6 \sigma^2 + 5 \sigma^4) \log(2 \sigma) 
    \nonumber
\\
&&
    + (-17 + 16 \sigma^2 - 5 \sigma^4) \log(1 - \sigma^2) 
    + 2 (4 - 5 \sigma^2 + 2 \sigma^4) \sin(2)
\Big] 
\nonumber
\\
&&
- 2 \sigma \left[
    (1 - \sigma^2)(7 - 5 \cos(2) + \sigma^2 (-7 + 3 \cos(2))) 
    + 2 (2 - 4 \sigma^2 + \sigma^4) \sin(2)
\right] \sin(2 \sigma)
\Big]\,.
\eea
\end{appendix}

{\small


\providecommand{\href}[2]{#2}\begingroup\raggedright\endgroup

}
\end{document}